\documentstyle[12pt,epsfig]{article}
\def\lapp{{\ \lower 0.6ex \hbox{$\buildrel<\over\sim$}\ }}
\def\gapp{{\ \lower 0.6ex \hbox{$\buildrel>\over\sim$}\ }}
\begin{document}
\begin{titlepage}
\vspace*{-1cm}
\begin{flushright}
FTUV/98/60 \\
IFIC/98/61   \\
July 1998 \\
\end{flushright}                                
\vskip 1.cm
\begin{center}                                                                  
{\Large\bf
Cascade decays of triplet Higgs bosons at LEP2.}
 \vskip 1.cm
{\large A.G.~Akeroyd\footnote{akeroyd@flamenco.ific.uv.es}}
\vskip .4cm
 Departamento de F\' \i sica Te\'orica, IFIC/CSIC,\\
 Universidad de Valencia, Burjassot 46100,\\
Valencia, Spain
\vskip 1cm                                                                    
\end{center}                                                                    

\begin{abstract}

We study the Georgi--Machacek two triplet, one doublet model in
the context of LEP2, and
show that cascade decays of Higgs bosons to lighter Higgs bosons
and a virtual vector boson may play a major role. 
Such decays would allow the Higgs bosons of this model
to escape current searches, and in particular are
of great importance for the members of the five--plet which will
always decay to the three--plet giving rise to cascade signatures.

\end{abstract}
\vfill
\end{titlepage}                                                                 
\newpage                                                                        

\section{Introduction}
It is well known that the Standard Model (SM) \cite{Wein}
 requires a Higgs sector \cite{Higgs} in order to 
break the electroweak symmetry and allow massive fermions
and gauge bosons.
 Complex Higgs doublets (and singlets) are the most 
natural way of achieving this because they predict
$\rho=M^2_W/(M^2_Z\cos^2\theta_W)=1$ at tree--level 
\cite{Gun}. 
With the present experimental value measured to be $1.0012\pm 0.0013\pm 
0.0018$ \cite{Rev}, such a tree--level result is preferable. 
The minimal SM possesses one complex  
doublet which after symmetry breaking predicts 
one physical neutral scalar ($\phi^0$), 
although this may not be nature's choice and much can be 
found in the literature concerning the Two--Higgs--Doublet model (2HDM) 
\cite{Gun}. Such extended Higgs sectors arise in many 
well--motivated extensions of the SM (e.g. Supersymmetric theories).

It may be that higher Higgs representations contribute to symmetry 
breaking and in this paper we consider an extended Higgs sector 
proposed by Georgi and Machacek which consists of
two Higgs triplets and a Higgs doublet $\cite{Geo}\to \cite{Chan}$.
This model (HTM) preserves $\rho=1$
at tree--level due to a cancellation between the triplet contributions.
 A fair amount of work has been done the HTM $\cite{Veg}\to 
 \cite{Bam}$, and in this paper we build upon the analysis of our
  earlier work \cite{Ake} which considered the phenomenology at LEP2.

Due to the large number of Higgs bosons in the HTM
its phenomenology may appear at first sight very complicated.
 However, we
shall see that the degeneracy among the various multiplets, and the 
constraint on the triplet field vacuum expectation value (VEV)
 simplify the phenomenology considerably and make the HTM
more predictive.
In particular, we show that the five--plet members are always heavier
than the three--plet members and that the former will always decay
dominantly to the
latter, giving rise to cascade signatures. For masses in 
the energy range at LEP2 these decays are often three--body, although
due to the fermiophobic nature of the five--plet they still dominate.  
In addition, we show that the decays of the three--plet bosons
to the eigenstate $\psi_2$ and a virtual vector boson are possible and
may be dominant.
The paper is organized as follows. In Section 2 we briefly introduce 
the HTM and show how precision measurements at LEP constrain its 
phenomenology. In Section 3 we evaluate limits
on the masses of the HTM scalars by using results from present
experimental searches for Higgs bosons. In Section 4 we calculate the 
branching ratios
(BRs) of the five--plet and three--plet members for masses relevant for LEP2,
and in Section 5 we examine the possible production processes and signatures
at this collider. Finally, Section 6 contains our conclusions.

\section{The Higgs Triplet Model (HTM)}
We do not attempt to give a detailed review of the HTM here and 
instead refer the reader to Refs.~\cite{Geo} and \cite{Veg}. 
The model possesses one isopin doublet (hypercharge Y=1) and 
two triplets with $Y=0$ and $Y=2$ respectively.
The particle spectrum consists of a degenerate
five--plet of Higgs bosons ($H_5^{\pm\pm}$, $H_5^{\pm}$, $H_5^0$),
a degenerate three--plet ($H_3^{\pm}$, $H_3^0$), and two singlets under
the custodial symmetry ($H_1^0$ and $H_1^0$$'$).
It is convenient to introduce a doublet--triplet mixing angle 
($0\le\theta_H\le \pi/2$) defined by \begin{equation} \cos{\theta}_H\equiv 
{a\over{\sqrt 
{a^2+8b^2}}}\;,\;\; {\sin{\theta}_H\equiv \sqrt{8b^2\over a^2+8b^2}}\;.
\end{equation}
Here $b$ is the VEV of the neutral triplet fields, and $a$ the VEV of the
neutral doublet field. The five--plet members and $H^{0}_1$$'$ are 
composed entirely of triplet fields, and so most tree--level couplings
 to fermions are forbidden by gauge invariance. The exception is the
 possibility of $H_5^{\pm\pm}$ being coupled to two leptons 
 (see Section 4.2). The 
three--plet members, $H^{\pm}_3$ and $H^0_3$, are respectively
 equivalent to $H^{\pm}$ 
and $A^0$ of the 2HDM (Model~I) with the replacement $\cot\beta\to 
\tan\theta_H$ in the Feynman rules. In the limit of $\tan\theta_H\to 0$
 (i.e. the triplet fields
do not contribute to symmetry breaking) $H^0_1$ plays the role
of $\phi^0$.
A full list of Feynman rules for the HTM appears in Ref.~\cite{Veg}.

Precision measurements of the process $Z\to b\overline b$ impose the 
strongest bound on $\sin\theta_H$. Virtual charged scalars with 
tree--level fermion couplings contribute to this decay \cite{Hollik},
\cite{Muk} e.g. $H^{\pm}$ in 
the 2HDM and $H^{\pm}_3$ in the HTM. Ref.~\cite{Giudice} shows that  
one can obtain the bound 
$\cot\beta \le 0.555$ ($95\%$ c.l) in the 2HDM for $M_{H^{\pm}}=85$ GeV, 
with the bound weakening slightly for larger $M_{H^{\pm}}$.
This corresponds to $\tan\theta_H<0.555$ in the HTM, 
improving the bound $\tan\theta_H<0.8$ that we quoted in Ref.~\cite{Ake}. 
Bounds from $B\overline B$ mixing can be competitive  
\cite{Chak}, although suffer from some uncertainties in the measured
values of the input parameters. 
Throughout the paper we shall be using $\tan\theta_H<0.555$, which is
justified since we are interested in the case of the three--plet being in
 range at LEP2.
We shall see that this
 constraint on $\tan\theta_H$ significantly effects the
 phenomenology of the HTM, making it more predictive.  

 We now consider the masses of the
Higgs bosons.
Mixing may take place between the fields $H^{0}_1$$'$ and $H_1^0$,
and there exist two mass eigenstates denoted by $\psi_1$ and $\psi_2$ 
with $M_{\psi_1}>M_{\psi_2}$. The mass matrix in the basis 
$H_1^0$--$H_1^0$$'$ is given by:
\begin{equation}
M=\left(\begin{array} {*{2}{c@{\;\;}}c@{\;\;}}
8c^2_H(\lambda_1+\lambda_3) & 2\sqrt 6 s_Hc_H{\lambda_3} \\
 2\sqrt 6 s_Hc_H{\lambda_3} & 3s^2_H({\lambda}_2+{\lambda}_3)
 \end{array} \right)v^2\;. 
  \end{equation}
The compositions of the mass eigenstates are: 
\begin{eqnarray}
 \psi_1=H^{0'}_1\sin\alpha+H^0_1\cos\alpha\;, \\
 \psi_2=H^{0'}_1\cos\alpha-H^0_1\sin\alpha\;. 
 \end{eqnarray}
We shall denote the common mass of the three--plet (five--plet)
 members as $M_3$($M_5$), with values given by:
\begin{equation}
M^2_3=\lambda_4v^2\;\;\;M^2_5=3(s^2_H\lambda_5+c^2_H\lambda_4)\;.
\end{equation}
Here $\lambda_i$ are parameters from the Higgs potential, $s_H=\sin\theta_H$
and $c_H=\cos\theta_H$.
If one allows $s^2_H\to 1$ (i.e. no constraint on $s_H$), 
which is often taken in the literature
to maximize the effects of exotic couplings that depend
 on $s_H^2$, there would exist a parameter space for $M_5\le M_3$. 
Assuming that the five--plet members are the lightest, Refs.~\cite{Veg},
$\cite{Ask}\to \cite{Ghosh}$ consider the decay channels of
$H^{0}_5$, $H^{\pm\pm}_5$ and $H^{\pm}_5$. They conclude that for $M_5$
in range at LEP2 one would find $H^{\pm\pm}_5\to W^{(*)}W^*$,
 $H^{\pm}_5\to
W^{(*)}Z^{*}$, and $H^{0}_5\to \gamma\gamma$ as the dominant decays.
 However, in the light of the bound on $s^2_H$ one finds
that the five--plet members are {\sl always} heavier than the 
three--plet members. 
Even equating $\lambda_5=0$ and $c_H^2$ at 0.764 (its lowest value)
 one would
find from Eq. (5) that $M_5\approx 1.5M_3$. This result has
important consequences for the 
phenomenology of the five--plet members, since decays to $H_3H_3$ 
or $H_3V$ ($V$ is $W$ or $Z$) will
always be available, with sometimes one of the particles off--shell. 
We will show that in the HTM  
these three--body decays to lighter Higgs bosons dominant the decays 
to two vector
bosons for $M_5$ in range at LEP2, and so would imply cascade
 decay signatures for the five--plet.\footnote{ 
The results of Refs.~\cite{Veg}$, \cite{Ask} \to \cite{Ghosh}$ will 
still apply to other models in which a fermiophobic
$H^{\pm\pm}$, $H^{\pm}$ or $H^0$ exists.}
A recent search at LEP2 for $H_5^0$ \cite{OPAL} assumed the decays 
 $H_5^0\to H_3V^{*}$ to be negligible and obtained the bound
 $M_5\ge 79.5$ GeV.

The masses of
$\psi_1$ and $\psi_2$ are dependent on $\lambda_1,\lambda_2,\lambda_3$,
and so there is no correlation between $M_{\psi_1}$,$M_{\psi_2}$
and $M_3$,$M_5$. Therefore one can consider the case of $\psi_2$
being lighter than $M_3$.
In Ref.~\cite{Ake} we showed that a natural argument of 
taking all $\lambda_i$ equal suggested $\psi_2$ to be
the lightest of the Higgs bosons.
It was also shown that very little mixing occurs and that one may take 
$H^0_1$ and $H^0_1$$'$ to be effectively mass eigenstates. 
With the improved bound $s^2_H\le 0.236$ one would find 
the mixing angle $\alpha$ constrained even more. 

We shall assume that $\psi_1$ is the heaviest of the Higgs bosons and is
 out of the LEP2 range -- the natural argument in Ref.~\cite{Ake}
 would suggest this. We shall consider two scenarios, bearing in
 mind that $M_3\le M_5$.
\begin{itemize}
\item [{(i)}]  $M_{\psi_2}\le M_3$

\item [{(ii)}] The three--plet members are the lightest of the Higgs bosons
\end{itemize}
These two situations may produce different experimental signatures, since
in case (i) we expect a parameter space 
for significant three body decays of the three--plet to $\psi_2$ 
(in an analogous way to the results in the 2HDM \cite{Ake3body}).
In case (ii) we expect the three--plet BRs to be identical
to those of the 2HDM (Model~I).

\section{Experimental limits on masses}

In this section we derive mass limits on the
 Higgs bosons of the HTM by using current experimental limits
valid for $\phi^0$ and the scalars of the 2HDM. First we consider $\psi_2$.
If one assumes $\psi_2=H^0_1$ (i.e. if $\cos\alpha=0$ in Eq. (4)) 
one may use the current
LEP searches for $\phi^0$ to place a direct mass bound since $H^0_1$ has 
essentially the same BRs as $\phi^0$ for masses in range at LEP; although
its 
fermion couplings are scaled by a factor $1/c_H$ and the vector boson
couplings by a factor $c_H$ relative to $\phi^0$,
since decays to the latter are
 negligible at LEP energies, one can apply the results from $\phi^0$ searches. 
Hence one obtains $H^0_1\ge 87.6$~GeV for $c^2_H=1$ \cite{L3}
(i.e. the $H^0_1ZZ$ coupling is $\phi^0ZZ$ strength). 
For the smallest value of $c_H$ ($c^2_H\ge 0.764$) one would find a limit
of $\approx 69.5$~GeV \cite{ALEPH}.  

If $\psi_2=H^0_1$$'$, (i.e. $\sin\alpha=0$)  $\psi_2$ would be
a fermiophobic Higgs ($H_F$) \cite{Diaz} and one would expect a very large
BR to $\gamma\gamma$. Using the production mechanism $e^+e^-\to H_FZ$, and
assuming that the cross--section is equal to that of $e^+e^-\to \phi^0Z$,
Ref.~\cite{OPAL} obtain the limit $M_F\ge 90$ GeV.
For $H^0_1$$'$ this limit may be weakened since the coupling
$H^{0}_1$$'$$ZZ$ (which determines the cross--section)
 is proportional to $8s^2_H/3$. 
For small $s_H$ a very light $H^0_1$$'$ could still be possible,
 and would be a 'hidden' fermiophobic Higgs. We note here that 
 Ref.~\cite{OPAL} also searches for $\gamma\gamma f\overline f$
  final states, {\sl without} requiring that the fermions originate from 
  a $Z$. Therefore this search is sensitive to the process
  $e^+e^-\to \psi_2H_3^0$, with $H^0_3\to f\overline f$. Ref.~\cite{OPAL}
 shows that the sum of the cross--sections must satisfy the following
  relation: 
 \begin{equation}
 \sigma(e^+e^-\to \psi_2Z)+\sigma(e^+e^-\to \psi_2H_3^0)\le 150\;{\rm fb}
 \end{equation}
 Since these two production processes are
 complementary, if one wishes to consider a light fermiophobic $\psi_2$
  Eq. (6) can only be satisfied if
 $e^+e^-\to \psi_2H_3^0$ is closed/suppressed -- i.e. $M_{\psi_2}+M_3\ge 160$
 GeV. A caveat here is that in these scenarios of a light $\psi_2$
 one would have a large parameter space for a significant/dominant 
 BR$(H_3^0\to Z^{(*)}\psi_2$), see Section 4.1. Although this would 
 produce a different event topology ($\gamma\gamma$ recoiling against
 $\gamma\gamma f\overline f$) much of the current selection criteria
 would still be relevant \cite{Turcot}. We conclude that a 
 light fermiophobic $\psi_2$ would most likely require
  $M_{\psi_2}+M_3\ge 160$ GeV,
 although $M_{\psi_2}+M_3\le 160$~GeV is not strictly ruled out.  
 
In the general case of $\psi_2$ being a mixture
of the above two states (i.e. if $\lambda_3\ne 0$ and $s_H\ne 0$),
one finds the coupling $ZZ\psi_2$ relative to that for $ZZ\phi^0$ to be:
\begin{equation}
ZZ\psi_2={2\sqrt 2\over \sqrt 3}\cos\alpha s_H-\sin\alpha c_H\;.
\end{equation}
The mixing angle $\alpha$ is found from:
\begin{equation} 
\sin2\alpha={2M_{12}\over \sqrt {(M_{11}-M_{22})^2+4M_{12}^2}}\;. 
\end{equation} 
Here $M_{ij}$ are the mass matrix entries in Eq. (2).
The angle $\alpha$ may take values from $-\pi/2\to \pi/2$
i.e. negative values if $M_{12}$ is negative ($\lambda_3<0$). 
$\lambda_3$ may be positive or negative, in contrast to $\lambda_4$ and
$\lambda_5$ which must be positive (\cite{Chan}).
 Ref.~\cite{Bam} plotted the coupling $ZZ\psi_2$ as a function of
$t^2_H$, allowing values of the latter up to 6.25.
We are interested in the region $t^2_H\le 0.31$
which is consequently not very clear in the graphs in Ref.~\cite{Bam}. 
In the light of the bound
on $t_H$, we find it more appealing to plot
the coupling in Eq. (7) as a function of $\alpha$, which was
 chosen to have fixed
values in Ref.~\cite{Bam}. We draw two
curves for $|ZZ\psi_2|^2$ corresponding to $t_H=0.555$ (maximum value) 
and $0.3$, and one can clearly see
not only the destructive interference for positive $\alpha$, but also the 
constructive interference which was not noticed in Ref.~\cite{Bam}. 
Thus in the general case of mixing,
$\psi_2$ may attain larger cross--sections than are possible for either 
of the individual fields
$H_1^0$$'$ and $H^0_1$. Such an enhancement is never possible in extended
models with only doublets, due to the familiar suppression factors
$\sin^2(\beta-\alpha)$ and $\cos^2(\beta-\alpha)$. 
In the case of destructive interference a very light $\psi_2$ is not
ruled out.  
Hence it is possible that $\psi_2$ could be light, 
and the presence of a light $\psi_2$ can drastically affect the
decay channels of the three--plet bosons (see Section 4.1).
\begin{figure}[hbt]
\centerline{\protect\hbox{\psfig{file=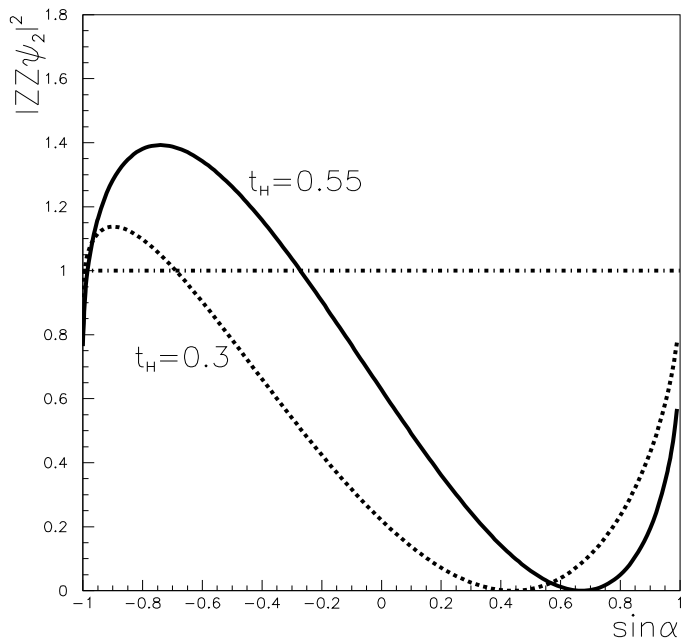,height=10cm,width=10cm}}}
\caption{Coupling $|ZZ\psi_2|^2$ relative to that of $|ZZ\phi^0|^2$ against
 $\sin\alpha$.}
\label{tri1}
\end{figure} 
 
In Fig.~2 we plot
BR$(\psi_2\to \gamma\gamma)$ as a function of $\sin\alpha$ for three
values of $t_H$. The decays of $\psi_2$ are dominated by the
component of $H^0_1$  unless $\sin\alpha$ is small which corresponds to
a large component of $H^0_1$$'$.
We recall that the natural scenario considered in Ref.~\cite{Ake} 
shows that
such tiny mixing would be expected, and that $\sin\alpha\le 0.05$ would
occur in the case of all $\lambda_i$ equal. 

\begin{figure}[hbt]
\centerline{\protect\hbox{\psfig{file=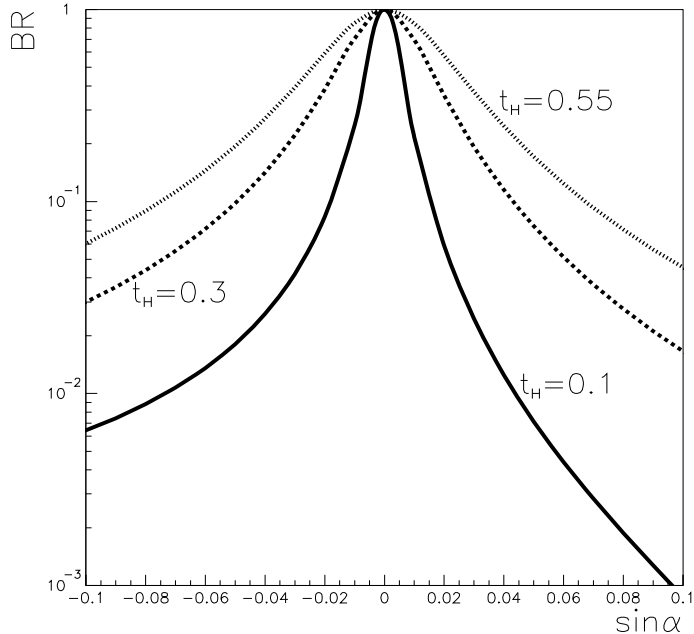,height=10cm,width=10cm}}}
\caption{BR$(\psi_2\to \gamma\gamma)$ against $\sin\alpha$, for $M_{\psi_2}=70$
GeV.}
\label{tri1}
\end{figure} 

For the three--plet bosons there are no direct mass limits on $H^0_3$
independent of $M_{\psi_2}$,  
in the same way that no bound on $A^0$ exists independent
of $M_{h^0}$ in the general 2HDM (apart from the bound $M_A\ge 5$~GeV from
considering the decay $\Upsilon\to A^0\gamma$ \cite{Ball}). In
the HTM one could use the bound $M_3+M_{\psi_2}\ge 90-110$ GeV
which is obtained from the combined search for $h^0$ and $A^0$ \cite{OPAL1}. 
A caveat here is that the cross--sections for 
$e^+e^-\to Z\psi_2$ and $e^+e^-\to H^0_3\psi_2$ may be larger than in 
2HDM (see Section 5) and so the bound on $M_3+M_{\psi_2}$ could be increased.
Since $H^0_3$ is degenerate
with $H^{\pm}_3$, any bound on the latter could be used for the former.
Current charged Higgs boson searches assume decays to $\tau\nu_{\tau}$ or
$cs$ and obtain the lower bound of 55~GeV ($95\%$ $c.l$) \cite{Cha}.
$H^{\pm}_3$ would decay in this way in the absence of a
light $\psi_2$. In the 2HDM (Model~I) the existence of a light neutral
Higgs 
can invalidate this limit, although this is not the case in 
HTM since the bound $M_3+M_{\psi_2}\ge 90-110$ GeV
also applies to $H^{\pm}_3$, thus ruling out the possibility of
$M_{\psi_2}\le M_3\le 55$ GeV.  

For the five--plet there again exist no direct limits since, as we shall 
show, these bosons would decay via cascades to the three--plet members, giving
final states that have not been searched for. However, since the five--plet is
heavier than the three--plet, one may obtain the bound $M_5\ge 82.5$ GeV, 
found by multiplying the bound on $M_3$ by 1.5. 

\section{Branching ratios}
In this section we study the branching ratios of the  five--plet members 
and the three--plet members.  The possible decays for the five--plet will
involve $H_5\to H_3H_3$ and $H_3V$, where one of the particles
may be off--shell for masses in range at LEP2 energies.
If $\psi_2$ were lighter than
the three--plet then the decay $H_3\to \psi_2V^*$ would be possible.  
Ref.~\cite{Veg} considers the decays of the five--plet to the
 three--plet, although for mass choices that are not relevant for LEP2 
i.e. they take $M_3=81$ GeV, which would give $M_5=121.5$ GeV, thus
taking the five--plet out of range. 
In addition, $t_H^2$ was fixed to the values of 0.01, 2.25 and 100, while
we wish to concentrate on the region $t_H^2\le 0.31$. 
We aim to give branching ratios for these
channels for masses in range at LEP2, making use of the 
fact that $M_5\ge 1.5M_3$ and so the vector boson is never very
off--shell in the decay $H_5\to H_3V^*$. In addition, Ref.~\cite{Veg}
did not consider the possibility of the decay $H_3\to \psi_2V^*$. 
 
In the following subsections we shall be neglecting the decays 
$H_5\to H_3^*V$ i.e. where the Higgs boson is off--shell 
and the vector boson is on--shell. Ref.~\cite{Veg} explains that these
are very small compared to $H_5\to H_3V^*$ since $V^*\to f\overline f$
is a gauge strength coupling while $H_3^*\to f\overline f$ is
Yukawa strength; moreover, the latter involves $t_H$ which can never
compensate for the smallness of the coupling. 

\subsection{The decays of the three--plet to $\psi_2$}
In this subsection we consider the decays $H^0_3\to Z^*\psi_2$
and $H^{\pm}_3\to W^*\psi_2$. Analogous decays
of $H^{\pm}$ and $A^0$ in the 2HDM (Model~I)
were shown to be dominant over a wide range of parameter space
in Ref.~\cite{Ake3body}. In the HTM, 
$\psi_2$ is a mixture of $H_1^0$ and $H_1^0$$'$ and the
width for this three--body decay is given by \cite{Djou},\cite{Arib}:
\begin{eqnarray}
\Gamma(H^{\pm}_3\to \psi_2W^*\to hf\overline f')={9G^2_FM^4_W\over 
16\pi^3}M_{H^\pm}CG_{\psi W}\;, \nonumber \\
\Gamma(H^{0}_3\to \psi_2Z^*\to hf\overline f)={6.1G^2_FM^4_W\over 
8\pi^3}M_{H^\pm}CG_{\psi Z}\;.
\end{eqnarray}
The analytical form of the function $G_{\psi V}$ may be used
for $M_3\ge M_W$. For $M_3\le M_W$ one must perform the integral
numerically. The coupling $C$ is given by
\begin{equation}  
C={8\over 3}\cos^2\alpha c^2_H+
\sin^2\alpha s^2_H+{4\sqrt 2\over \sqrt 3}s_Hc_H\sin\alpha\cos\alpha\;. 
\end{equation}

\begin{figure}[hbt]
\centerline{\protect\hbox{\psfig{file=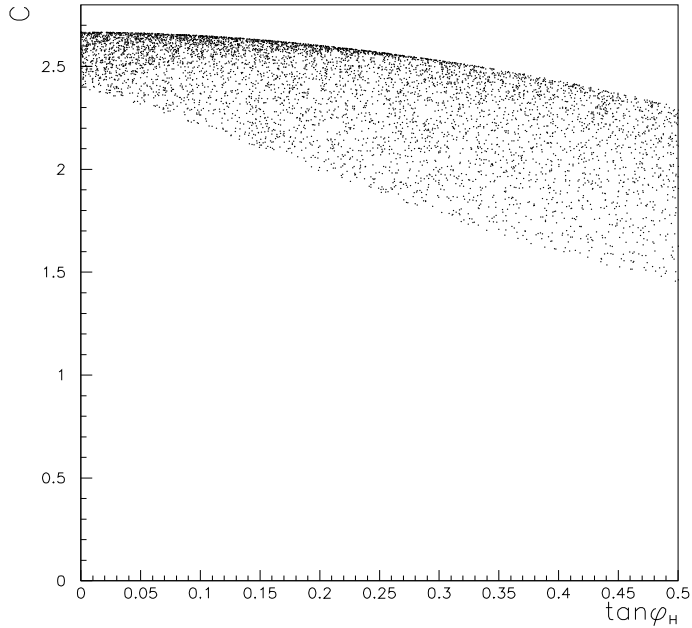,height=10cm,width=10cm}}}
\caption{The coupling $C$ as a function of $t_H$, demanding a light $\psi_2$.}
\label{tri1}
\end{figure} 
In order to allow a very light $\psi_2$ which has escaped
detection at LEP one must impose a condition on the 
$ZZ\psi_2$ coupling (Eq. (7)). For example, for $M_{\psi_2}\le 34$~GeV one 
requires $\sigma(e^+e^-\to Z^*\psi_2)\le 0.1
\sigma(e^+e^-\to Z^*\phi^0)$ \cite{ALEPH}.
Imposing this condition we plot in Fig.~3 the value of $C$ 
as a function of $t_H$ for 5000 random 
values of $\alpha$ and $\theta$. 
We see that smaller $t_H$ causes larger $C$,
and so decreasing $t_H$ enhances the three--body width and simultaneously
reduces the widths of the competing decays, since 
$\Gamma(H_3\to f\overline f$) is proportional to $t^2_H$. 
We also note that the value of $C$ is always greater than 1, and so the
 relative strength
of these three--body decays is greater than those for the analogous 
decays in the
2HDM (Model~I) in which $C=1$ or $\cos^2(\beta-\alpha$). Hence in order
to see their magnitude in the HTM it is sufficient to use the figures in
Ref.~\cite{Ake3body} bearing in mind that the widths in the HTM will
be greater by a factor $C$ i.e. use Figs.~2,3 for $H_3^{\pm}$ and Fig.~7
for $H^0_3$ with appropriate scaling and interpreting $\tan\beta$ as 
$\cot{\theta_H}$.   
We note that these decays of the three--plet bosons
were not considered in Ref.~\cite{Veg}, although
may be dominant or even close to $100\%$ over a wide 
range of parameter space.
In addition, they would be an alternative way of producing a very light
$\psi_2$ which is escaping current searches.

\subsection{The decays of $H_5^{\pm\pm}$}
The decays we shall consider are the following:
\begin{itemize}
\item [{(i)}] $H_5^{\pm\pm}\to WW^*$   

\item [{(ii)}] $H_5^{\pm\pm}\to H_3^{\pm}W^{*}$

\item [{(iii)}] $H_5^{\pm\pm}\to H_3^{\pm}H_3^{\pm}$
\end{itemize}
There is another possible decay, that is, $H_5^{\pm\pm}\to l^{\pm}l^{\pm}$
 (for recent work 
see Refs.~\cite{Chou}, \cite{Gunlep}).
In the HTM these decays are only significant if $s_H$ is very small,
since the width is given by
\begin{equation}
\Gamma(H_5^{\pm\pm}\to l^{\pm}l^{\pm})={|h_{ll}|^2\over 8\pi}M_5\;.
\end{equation}
\begin{figure}[hbt]
\centerline{\protect\hbox{\psfig{file=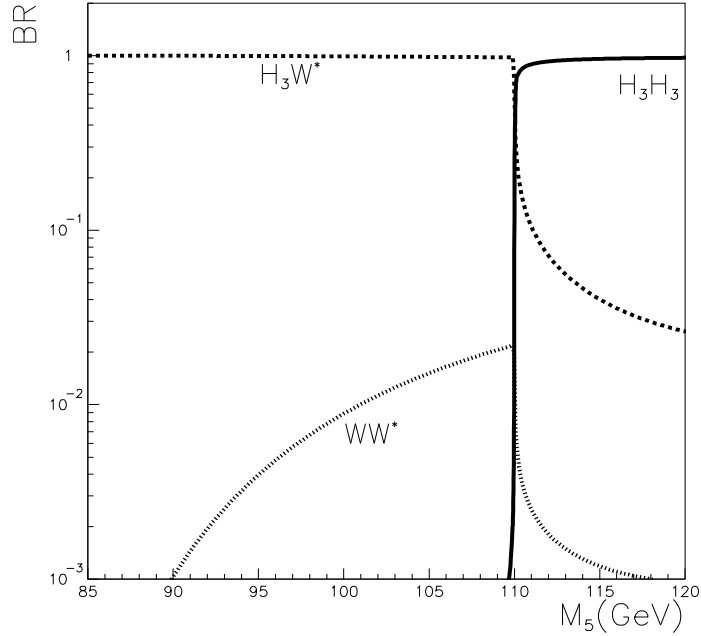,height=10cm,width=10cm}}}
\caption{BRs of $H_5^{\pm\pm}$ with $\tan\theta_H=0.555$.}
\label{tri1}
\end{figure}

The current neutrino mass limits constrain $h_{\tau\tau}\le 
1.4\times 10^{-4}/s_H$, while the couplings $h_{\mu\mu}$ and  $h_{ee}$
are smaller by 3 orders and 8 orders of magnitude respectively.
Ref.~\cite{Chou} considered the decays of an $H^{\pm\pm}$ in a model
with a Higgs doublet and only one triplet. They considered decays
(i) and (ii) and the bi--lepton channels are always strong since
$s_H\le 0.0056$ ($95\%$ $c.l$) in this model. We shall neglect the 
bi--lepton channels since in the HTM since one usually considers larger
values of $s_H$.  Ref.~\cite{Chou} 
found that the decay (ii) is important,
and we find the same. Importantly, the HTM {\sl requires} $M_5\ge M_3$
and so this channel is always open, and we also expect the bi--lepton
channel to be small.
In Fig.~3 we plot the BRs for the channels (i), (ii) and (iii)
as a function of $M_5$, fixing $M_3=55$~GeV (i.e. its lower limit) and
$t_H=0.555$.
 We see that the three--body decay (ii) is close to $100\%$ until the real
threshold for $H_3H_3$ decays is reached. Beyond this threshold the
decay (iii) dominates. We note that the
decay to two vector bosons does not surpass a BR of $2\%$. 
The domination of channel (ii) before the
real $H_3H_3$ threshold is reached is due to two reasons;
the decay is not severely phase space suppressed since $M_5\ge 1.5M_3$,
and the coupling $H_5^{\pm\pm}H^{\pm}_3W$ is proportional to $c_H$.
The off--shell decay  $H_3H_3^{*}$ is included in our plots although
is small. Lowering the value of $t_H$ causes the decay (iii) to
reach $100\%$ more quickly, and further reduces decay (i).

\subsection{The decays of $H_5^{\pm}$}
The decays we shall consider are the following:
\begin{itemize}
\item [{(i)}] $H_5^{\pm}\to Z^*W^{(*)}$   

\item [{(ii)}] $H_5^{\pm}\to H_3^{\pm}Z^*$

\item [{(iii)}] $H_5^{\pm}\to H_3^{0}W^*$

\item [{(iv)}] $H_5^{\pm}\to H_3^{0}H_3^{\pm}$

\end{itemize} 
\begin{figure}[hbt]
\centerline{\protect\hbox{\psfig{file=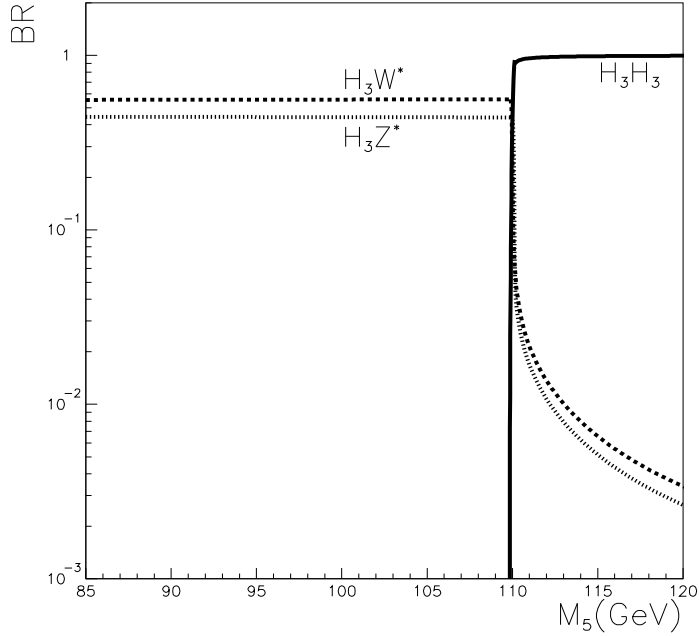,height=10cm,width=10cm}}}
\caption{Same as Fig.~4 but for $H_5^{\pm}$.}
\label{tri1}
\end{figure}

Here there are two possible decays to a three--plet member
and a vector boson, (ii) and (iii),
and decay (i) is not possible in models with only Higgs
doublets. The width of decay (i) was calculated in Ref.~\cite{God}
 although here we will see that this channel has a much smaller 
width than the decays (ii) and (iii).  In Fig.~4 we plot
the analogy of Fig.~3 for $H_5^{\pm}$. Decay (i), which would be
the dominate channel if $H_5^{\pm}$ were the lightest, is not included
in our plots although would peak at BR$\approx 0.3\%$ in the region
just before $M_5=110$ GeV. Again the three--body decays
to the three--plet and a virtual vector boson dominate until the
real $H_3H_3$ threshold is reached. We see that the decay 
mediated via $W^*$ is stronger than that mediated via $Z^*$. 
This is due to the fact that for a given $M_5$ and $M_3$, $Z$ is more 
off--shell than $W$, which compensates for the slightly weaker couplings
of the $W$ mediated decay in Eq.~(9).

\subsection{The decays of $H^0_5$}
The decays we shall consider are the following:
\begin{itemize}
\item [{(i)}] $H_5^0\to H_3^{\pm}H_3^{\mp}$   

\item [{(ii)}] $H_5^0\to H_3^0H_3^0$

\item [{(iii)}] $H_5^0\to H_3^{\pm}W^*$

\item [{(iv)}] $H_5^0\to H_3^0Z^*$

\item [{(v)}] $H_5^0\to \gamma\gamma,W^*W^{(*)},Z^*Z^{(*)}$

\end{itemize} 
\begin{figure}[hbt]
\centerline{\protect\hbox{\psfig{file=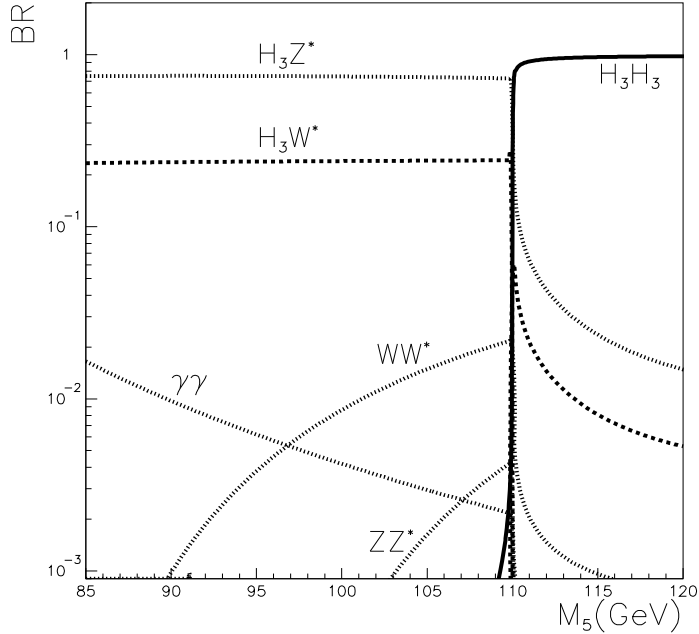,height=10cm,width=10cm}}}
\caption{Same as Fig.~4 but for $H_5^0$.}
\label{tri1}
\end{figure}
In the absence of decays to the three--plet $H_5^0$ would decay
predominantly (for $M_5\le 90$ GeV) to 
a mixture of $\gamma\gamma$ and $f\overline f$ 1--loop
induced decays \cite{Veg}. In Fig.~5 we plot the analogy of 
Fig.~3 for $H_5^0$. Again, decays (iii) and (iv) share the domination until
the real threshold for decays (i) and (ii) is reached. We note that
the $Z^*$ mediated decay is stronger than that mediated by $W^*$, 
in contrast to case
for $H_5^{\pm}$; this is due to the fact that the coupling $H_5^0H_3^0Z$
contains an extra factor of two with respect to the coupling
$H_5^0\to H_3^{\pm}W$. In the region $M_5\ge 2M_3$, decay (ii) has partial
width four times greater than that of decay (i). The BR to $\gamma\gamma$
is reduced to values less than $2\%$, and this result would effect
current search techniques \cite{OPAL} which only assume decay channels (v).

\section{Production Channels at LEP2}
In this section we consider the possible signatures of the Higgs bosons
of the HTM at LEP2. Some of the production channels are analogies of 
production channels in models with only Higgs doublets, while others
 are particular to models
with higher representations. In the charged Higgs sector
one has:
\begin{equation}
e^+e^-\to H_3^+H_3^-\;, H_5^+H_5^-\;,H_5^{++}H_5^{--}\;, 
H_3^+H_5^-\;, W^{\pm}H_5^{\mp}\;.
\end{equation}
Identical pair production (the first three) has received a lot of coverage
 in the literature
\cite{Veg},\cite{Kom}, while the last two are not possible in models 
with only doublets. The mechanism $e^+e^-\to W^{\mp}H_5^{\pm}$ has 
received some
attention in the literature \cite{God},\cite{Phil} although
suffers from the suppression factor $s^2_H$. Non--identical pair
production 
can be sizeable if kinematically allowed, being
proportional to $c^2_H$ and having a permutation factor of 2.
In addition, it can be open when the pair production of the 
five--plet bosons is not.
In the neutral sector one has the bremsstrahlung channels:
\begin{equation}
e^+e^-\to ZH^0_5\;,Z\psi_2
\end{equation}
and the pair production channels: 
\begin{equation}
e^+e^-\to H^0_3H^0_5, H^0_3\psi_2\;. 
\end{equation}
The phenomenology of the three--plet would look very
similar to that of the 
2HDM (Model~I) since $H^0_3$ and $H^{\pm}_3$
have the same couplings as $A^0$ and $H^{\pm}$ respectively, with the
replacement $\cot\beta\to \tan\theta_H$. If there exists a 
light $\psi_2$ then one may use the analysis in Ref.~\cite{Ake3body}, 
bearing
in mind the decays of the three--plet to $\psi_2$ (Section 4.1)
 may be stronger than
the analogous decay in the 2HDM. In addition, the mechanism $e^+e^-\to 
\psi_2H^0_3$, being proportional to $C$ in Eq. (9), can have a 
larger cross--section than that for $e^+e^-\to A^0h$ in the 2HDM;
this is also the case for $e^+e^-\to Z\psi_2$ (see Fig.~1). 

If the five--plet is in range one would expect high--multiplicity cascade
signatures. In Table 1 we outline the possible signatures
of the five--plet members from the various
production mechanisms. We assume $\sqrt s=192$~GeV and the entries in
brackets allow the additional cascade decay $H_3\to \psi_2V^*$; 
'not open' indicates
that the production channel in question would not be kinematically allowed. 
\begin{table}[htb]
\centering
\begin{tabular} {|c|c|c|} \hline
 & $H_5\to H_3V^*$  & $H_5\to H_3H_3$   \\ \hline
$H_5^{++}H_5^{--}$ & 8f (12f) & not open \\ \hline
$H_5^+H_5^-$ & 8f (12f) & not open \\ \hline
$H_5^0Z$ & 6f (8f)  & not open \\ \hline
$H_5^{\pm}W$ & 6f (8f) & 6f (10f)  \\ \hline 
$H_5^{\pm}H_3^{\mp}$ & 6f (10f)  &  6f (12f) \\ \hline 
$H_5^0H_3^0$ & 6f (10f) & 6f (12f) \\ \hline
\end{tabular}
\caption{The signatures of the five--plet members.}
\end{table}

The combination of the constraints $M_5\ge 1.5M_3$ and $M_{\psi_2}+
M_3\ge 110$ GeV
means that there is only a small parameter space open at LEP2 which
allows the cascade decay of $H_5\to \psi_2X$ via the three--plet 
to be open. For example, mass choices
such as $M_5=90$~GeV, $M_3=60$~GeV and $M_{\psi_2}=50$~GeV would
allow pair production of the five--plet to be open and satisfy the
mass relations above; in these cases the difference
between $M_3$ and $M_{\psi_2}$
is small and so one would need smaller $\tan\theta_H$ in order to
allow a substantial BR for the three--body decay $H_3\to \psi_2V^*$.  
We note that for single production of $H_5^{\pm}$, $H^0_5$ in association
with a gauge boson one could find asymmetric topologies such as
2 fermions recoiling against 6 or 8 fermions. For the signatures
in the first column one can always trigger 
on leptons originating from the virtual vector bosons. Jets originating from
$H_3$ could be reconstructed to give the mass of the three--plet bosons.
We conclude that the signature of the five--plet would be a large
multiplicity fermion event.

\section{Conclusions}
We have studied the phenomenology of the 2 triplet, 1 doublet model (HTM)
in the context of LEP2. We showed that current precision measurements
tightly constrain the triplet--doublet VEV ratio, and so cause the five--plet
($H_5$) bosons to be heavier than the three--plet $(H_3)$. We found that
the BR($H_5\to H_3V^*)\approx 100\%$ until the real threshold
for $H_5\to H_3H_3$ decays is reached. This ensures
that the signature of the five--plet bosons (if in range at LEP2)
 would be a large multiplicity fermion event.  We also showed that 
 $\psi_2$ may possess a production cross--section at LEP2 up to
 1.4 times that of the minimal standard model Higgs. Such an enhancement
 is never possible in models with only doublet representations. Conversely,
 $\psi_2$  may also be very weakly coupled to the $Z$ which would allow
 a light $\psi_2$ to have eluded detection at LEP. In this case the decays 
 $H_3\to \psi_2V^*$ are allowed and we showed that they may be the dominant
 channel over a wide region of parameter space, a result that
 would affect current search techniques for the three--plet and be an
  alternative way of searching for a weakly coupled $\psi_2$.

\section*{Acknowledgements}
This work was supported by DGICYT under grants PB95-1077, by the
TMR network grant ERBFMRXCT960090 of the European Union, and by a
CSIC--UK Royal Society fellowship. I wish to thank A. Turcot for 
useful comments and for proofreading the article.


\begin{thebibliography}{99}
\bibitem{Arib} I wish to thank A. Arhrib for
pointing out that the formula displayed in Ref.~28 for 
$\Gamma(H^\pm\to W^*A(h))$
requires a extra multiplicative factor of 0.5.
\end{thebibliography}
\end{document}